\documentclass[twocolumn,aps,superscriptaddress,nofootinbib,floatfix]{revtex4-2}
\usepackage{amsfonts}
\usepackage{amsmath}
\usepackage{float}
\usepackage{placeins}
\usepackage{graphicx}
\usepackage[hidelinks]{hyperref}
\usepackage{color,amsmath,amssymb,bm}
\usepackage{tensor}
\usepackage{lipsum}

\usepackage[normalem]{ulem}  

\renewcommand{\sout}{\bgroup \color{black} \ULdepth=-.5ex \ULset}

\def\blfootnote{\xdef\@thefnmark{}\@footnotetext}

\newcommand{\beq}{\begin{equation}}
\newcommand{\eeq}{\end{equation}}
\newcommand{\bea}{\begin{eqnarray}}
\newcommand{\eea}{\end{eqnarray}}

\begin{document}

\title{Anisotropic 
fluctuations of momentum and angular momentum
of heavy quarks in the pre-equilibrium stage of pA collisions
at the LHC}

\author{Gabriele Parisi}
\email{gabriele.parisi@dfa.unict.it}
\affiliation{Department of Physics and Astronomy, University of Catania, Via S. Sofia 64, 1-95125 Catania, Italy}
\affiliation{Laboratori Nazionali del Sud, INFN-LNS, Via S. Sofia 62, I-95123 Catania, Italy}

\author{Vincenzo Greco}
\email{vincenzo.greco@dfa.unict.it}
\affiliation{Department of Physics and Astronomy, University of Catania, Via S. Sofia 64, 1-95125 Catania, Italy}
\affiliation{Laboratori Nazionali del Sud, INFN-LNS, Via S. Sofia 62, I-95123 Catania, Italy}

\author{Marco Ruggieri}
\email{marco.ruggieri@dfa.unict.it}
\affiliation{Department of Physics and Astronomy, University of Catania, Via S. Sofia 64, 1-95125 Catania, Italy}
\affiliation{INFN-Sezione di Catania, Via S. Sofia 64, I-95123 Catania, Italy}

\date{\today}

\begin{abstract}
We simulate the real-time evolution of the SU(3)--glasma generated in the early stages of high-energy proton–nucleus collisions, employing classical lattice gauge theory techniques. Our setup incorporates a realistic modeling of the proton’s internal structure and includes longitudinal fluctuations in the initial state, enabling the study of genuinely non-boost-invariant collision dynamics. Focusing on the momentum and angular momentum anisotropies of heavy quarks in the infinite mass limit, 
we find that the system retains significant anisotropy well beyond the characteristic timescale $\tau = 1/Q_s$, being $Q_s$ the saturation energy of the glasma, the only scale in our model. This persistence of anisotropy is further confirmed in the more realistic, non-boost-invariant scenario, across a range of fluctuation amplitudes. These findings pave the way for future investigations involving dynamical heavy quarks and more quantitative initializations of the glasma.

\end{abstract}

\maketitle

\section{Introduction}

The analysis of a relativistic heavy-ion collision, as performed at the Relativistic Heavy Ion Collider (RHIC) at Brookhaven National Laboratory and at the Large Hadron Collider (LHC) at CERN, provides a unique opportunity to investigate the behavior of QCD matter at extreme temperatures and densities. The study of such events requires different kinds of descriptions of the spacetime evolution of the system, since the matter produced in the collision experiences different stages in its evolution. For instance, over the past few decades many experimental results have indicated the emergence of a new state of matter, referred to as the Quark-Gluon Plasma (QGP), which forms after $\sim 1$ fm after the collision and in which quarks and gluons behave as a hydrodynamic fluid \cite{Heinz:2000bk, Heinz:2013th, Harris:1996zx}.

If we instead consider the early stages of the collision ($\leq 1$ fm), the Classical Yang-Mills (CYM) theory offers a good description of this phase. It can describe well the nonequilibrium evolution of the highly occupied gluonic system, called the glasma, that appears immediately after the collision. In particular, the glasma simulation with the CYM field plays an important role in understanding the nonequilibrium stage between the moment of the collision and the onset of the hydrodynamic evolution of the QGP. In fact the glasma simulation, in the analysis of experimental data, is widely used to establish the initial conditions for the subsequent hydrodynamic evolution \cite{Schenke:2012hg,Schenke:2012wb}. The theoretical background for why the CYM theory offers a good picture of the initial gluonic matter is based on the  Color Glass Condensate  (CGC) framework, which stands as a valid description of the high-energy nucleus \cite{McLerran:1993ni,McLerran:1993ka,McLerran:1994vd}, see \cite{Gelis:2010nm,Iancu:2003xm,McLerran:2008es,Gelis:2012ri} for reviews. In such a high-energy nucleus the dominant degrees of freedom are soft gluons emitted from hard partons. In particular, the McLerran-Venugopalan (MV) model in the CGC effective theory describes the soft gluons as the CYM fields and the hard partons as their color sources. Consequently, the glasma generated in the collision of such high-energy nuclei can be described well by the CYM field. The greatest simplicity of the CGC formalism is the fact that the complexity of a non-linear many-body problem at high energy is reduced into a one-scale problem, with a hard saturation scale $Q_s$ as the only dimensionful relevant scale at which gluon recombination effects start to become as important as the gluon radiation ones. The CYM equations can be studied both analytically (e.g. by employing a small time expansion \cite{Carrington:2020ssh,Carrington:2021qvi}) or numerically \cite{Avramescu:2023qvv, Avramescu:2024poa}.

However, such a success of the glasma simulations has been largely limited by the boost invariance assumption, which shows good agreement with experimental data only around the midrapidity region. Recently, much attention has been paid to the 3 + 1-dimensional (3 + 1D) glasma simulations,
which aim to go beyond the boost invariance assumption, as necessary to understand observables across a broader region of rapidity \cite{Pang:2012uw,Pang:2014pxa}. In fact, on top of the glasma it is possible to add quantum
fluctuations \cite{Fukushima:2013dma, Romatschke:2005pm, Romatschke:2006nk, Fukushima:2011nq, Iida:2014wea, Epelbaum:2013ekf, Epelbaum:2013waa, Ryblewski:2013eja, Ruggieri:2015yea,Tanji:2011di, Berges:2012cj, Berges:2013fga, Berges:2013lsa, Berges:2013eia},  which are known to trigger plasma
instabilities \cite{Bazak:2023kol} and are helpful to produce entropy during
the early stages of high energy nuclear collisions \cite{Iida:2014wea, Tsukiji:2016krj,Tsukiji:2017pjx,Matsuda:2022hok}. Quantum fluctuations appear when one considers the finite
coupling corrections to the glasma solution (which is obtained in the small coupling limit): the spectrum of these
fluctuations has been computed within a perturbative
calculation in \cite{Epelbaum:2013ekf} and it has been shown that they affect
both the gauge potential and the color electric field.

The experimental framework we mentioned refers to Pb-Pb (LHC) or Au-Au (RHIC) collisions. Moving on to measurements in smaller collision systems such as pp and pA \cite{Dusling:2015gta,Greif:2019ygb} (in particular those of anisotropies in multi-particle correlation functions), these have shown very similar features as those found in heavy ion collisions. While calculations within the hydrodynamic framework have been quite successful in describing observables in these small collision systems, alternative explanations relying entirely on intrinsic momentum correlations of the produced particles can also reproduce many features of the experimental data \cite{Dusling:2015gta,Greif:2019ygb}. It is likely that both initial intrinsic momentum correlations and a hydro-like anisotropic expansion are necessary to understand the elliptic anisotropic emission in pA \cite{Greif:2017bnr, Nugara:2024net}. However, regardless the existence of alternative explanations, the applicability of hydrodynamics becomes increasingly doubtful as the system size decreases and gradients increase \cite{Kurkela:2018qeb,Kurkela:2019kip}. For this reason, a thorough treatment of the initial stages is even more important for pA collisions.
In this work we will simulate a pA collision by solving the classical Yang-Mills equations in SU$(3)$ with MV initial conditions for the initial charge. The original MV hypothesis will be extended in order to account for the quark structure of the proton.

Particularly sensitive probes of the very early stages of the collision are heavy quarks \cite{Dong:2019unq}. Due to their short formation time (of the order of $1/2M$, where $M$ is the heavy quark mass), they experience the initial stage of the collision. Moreover, their mass is greater than $\Lambda_{\text{QCD}}$, hence their initial production can be studied with perturbative methods. By understanding the imprint of the glasma fields on these probes, one can disentangle important information about the structure of initially produced matter, in both proton-nucleus and nucleus-nucleus collisions. There exist a relevant limiting case in which the accumulated momentum of hard probes in glasma can be evaluated only from glasma lattice field correlators \cite{Boguslavski:2020tqz,Avramescu:2023qvv}, without explicitly solving the particle equations of motion: this corresponds to infinitely massive heavy quarks.

The paper will be structured as follows: in Section \ref{The evolving Glasma} we will elaborate on the formalism used for solving the CYM equations, starting from the charge generation, then moving on towards the initialization of the color-magnetic fields and their time evolution. In Section \ref{Quarks of infinite mass in Glasma} we deal with the Wong equations governing the motion of Heavy Quarks (HQs), which are charm and beauty (the top quark here is not taken into account, since its huge mass implies decay times much smaller than the QCD timescales, check e.g. \cite{Apolinario:2017sob} for a study on top quarks in QGP) to find relevant quantities like the momentum shift and the angular momentum shift of infinitely massive HQs in glasma. In Section \ref{Initial state fluctuations} we aim to generalize the CYM formalism including initial state fluctuations, which break boost invariance. After showing the results, we then summarize and conclude.

In this article we use natural units $k_B=c=\hbar=1$. Moreover, throughout the paper we will use both Minkowski and Milne coordinates. The change of coordinates from one system to the other is given by
\begin{equation}
    \tau=\sqrt{t^2-z^2}, ~~~~~ \eta=\frac12 \log \frac{t+z}{t-z},
    \label{15.2}
\end{equation}
which give a definition of proper time and spatial rapidity.

\section{The evolving glasma}
\label{The evolving Glasma}

Several different strategies have been used to understand and describe the earliest phase of relativistic heavy-ion collisions. One which is commonly applied is the Color Glass Condensate effective theory \cite{McLerran:1993ni,McLerran:1993ka,McLerran:1994vd}, which is based on a separation of scales between the high Bjorken-$x$ degrees of freedom, i.e. the valence partons, and the low $x$ degrees of freedom which they generate, that is soft gluons.

\subsection{Color charges for p and A}
In our work, to realize the aforementioned separation,
the sources of the initial gluon fields 
are generated using the McLerran-Venugopalan (MV) model, 
in which the large-$x$ color sources 
are randomly distributed
on an infinitely thin color sheet.
The distribution of these charges is a gaussian, characterized by
zero average 
\begin{equation}
    \langle \rho^a (\mathbf{x}_\perp)\rangle=0,
    \label{1}
\end{equation}
and variance given by
\begin{equation}
    \langle \rho^a (\mathbf{x}_\perp)\rho^b (\mathbf{y}_\perp)\rangle=g^2 \mu^2 \delta^{ab} \delta^{(2)}(\mathbf{x}_\perp-\mathbf{y}_\perp).
    \label{2}
\end{equation}
Here, $g$ is the coupling constant (we use $g=2$,
corresponding to $\alpha_s=0.3$), 
and $\mu$ is the so-called MV parameter, describing the number density of the color charges per unit
of area in the transverse plane. The symbol $\perp$ here and throughout the paper refers to quantities in the transverse plane.
In our implementation, we relax the single-sheet hypothesis of the 
original MV model 
by generating a certain number $N_s$ of color sheets stacked on top of one another \cite{Lappi:2007ku}. 
Consequently, for each of the sheets,
the numerical implementation of Eq.~\eqref{2}
we will use in this work is
\begin{equation}
    \langle \rho^a_{n,x}\rho^b_{m,y}\rangle=g^2 \mu^2 \frac{\delta_{n,m}}{N_s}\delta^{a,b} \frac{\delta_{x,y}}{a_\perp^2},
    \label{3}
\end{equation}
where $a,b=1,\dots, N_c^2-1$ are SU(3) color indexes, $n,m=1,\dots,N_s$ are the color sheet indexes and $x,y$ span each point of a $N_\perp\times N_\perp$ lattice, whose transverse length is $L_\perp$ and lattice spacing is $a_\perp=L_\perp/N_\perp$. Operatively, the conditions \eqref{1} and \eqref{3} are satisfied by generating random Gaussian numbers with mean zero and standard deviation equal to $\sqrt{(g^2\mu^2)/(N_s a_\perp^2)}$: for a gaussian distribution of mean zero and standard deviation $\sigma$, one has $\langle x\rangle=0$ and $\langle x^2 \rangle=\sigma^2-\langle x\rangle^2=\sigma^2$, moreover the appearance of the factor $a_\perp$ comes from the discretization of the Dirac delta.
Below we explain how to build up the gluon fields generated
by each sheet, and how to
combine all these fields to obtain the initial
gluon field in the glasma.
We will use $N_s=50$ in Eq.~\eqref{3}
throughout this work: we have seen that there is a slight dependence of the energy density on the number of color sheets, which stops around this value of $N_s$. For a numerical evidence of this fact, check e.g. the red line in Figure 4 of \cite{Lappi:2007ku}, in which we see convergence of the extracted value of $Q_s$ at around $N_s\sim 50$: the value $N_s=50$ therefore represents a good compromise between the computational cost and the independence of observables from $N_s$ at larger values.
Moreover, such a value smears the statistical fluctuations due to the aforementioned random charge extraction.

These steps are shared by the charge generation of both a proton and a heavy ion. Indeed, the original MV model assumed a source charge density that was homogeneous in the transverse plane, which is feasible when dealing with large-A nuclei, of which we can neglect the details at the level of single nucleons. On the other hand, a reasonable description of protons involves a varying nuclear density in the transverse plane, in order to take into account for the underlying quark structure. The main difference between those two cases lies in the choice of $\mu$: in the A-case $\mu$ is chosen as a constant equal to $\mu=0.5$ GeV, which translates into uniform fluctuations of color charge throughout the lattice. The p-case is instead more involved. Let us introduce the the thickness function of the proton, $T_p$, as
\begin{equation}
T_p(\mathbf{x}_\perp) = 
\frac{1}{3} \sum_{i=1}^3 \frac{1}{2\pi B_q} \exp\left(-\frac{(\mathbf{x}_\perp-\mathbf{x}_i)^2}{2B_q}\right),
\label{eq:bd1}
\end{equation}
where the $\mathbf{x}_i$ denote the positions of the constituent quarks, which are randomly extracted from the distribution
\begin{equation}
T_{cq}(\mathbf{x}_\perp) = \frac{1}{2\pi B_{cq}}
\exp\left(-\frac{\mathbf{x}_\perp^2}{2B_{cq}}\right).
\label{eq:bd2}
\end{equation}
Parameters have been chosen as $B_q=0.3$ GeV$^{-2}$, $B_{cq}=4$ GeV$^{-2}$, i.e. the widths of the two gaussians differ by a factor around 4. After doing so, we evaluate the saturation scale $Q_s$ in this model as
\begin{equation}
Q_s^2(x,\mathbf{x}_\perp) = \frac{2\pi^2}{N_c}  \alpha_s\cdot 
xg(x,Q_0^2)\cdot T_p(\mathbf{x}_\perp).
\label{eq:qs_2_77}
\end{equation}
from which $\mu(x,\mathbf{x}_\perp)$ of the proton is given by \cite{Schenke:2020mbo}
\begin{equation}
g^2\mu(x,\mathbf{x}_\perp) = c Q_s(x,\mathbf{x}_\perp),
\label{eq:gmu_ajk}
\end{equation}
with $c=1.25$. Note that by virtue of Eq.~\eqref{eq:qs_2_77} not only $\mu$, but also $Q_s$ depend on the transverse plane coordinates.
Also, in principle $xg$ in \eqref{eq:qs_2_77} should be computed at the scale $Q_s$ by means of the Dokshitzer–Gribov–Lipatov–Altarelli–Parisi (DGLAP) equation with a proper initialization. This was done in \cite{Schenke:2020mbo,Rezaeian:2012ji}, and we reserve this approach for future studies. For the sake of simplicity, in the present study we limit ourselves by assuming that $xg$ is given by the initial condition at 
$Q_0^2=1.51$ GeV$^2$ as in \cite{Schenke:2020mbo,Rezaeian:2012ji}, namely
\begin{equation}
xg(x,Q_0^2)=A_g x^{-\lambda_g}(1-x)^{f_g},
\label{eq:anna23}
\end{equation}
with $A_g=2.308$, $\lambda_g=0.058$ and $f_g=5.6$. This simplification is partly justified by the fact that the average saturation scale for the proton is of the order of $Q_s\sim 1$ GeV, hence we do not expect the DGLAP evolution of the $xg$ in Eq.~\eqref{eq:anna23} to lead to significant changes. For pA collisions at the LHC energy, the relevant values of $x$ are in the range $[10^{-4},10^{-3}]$ {\cite{Schenke:2012hg,Schenke:2020mbo}. In our calculations, we will thus fix $x$ in the aforementioned range, then compute $xg$ by virtue of Eq.~\eqref{eq:anna23}. If not otherwise stated, we will use the corresponding value obtained for $x=10^{-4}$, that is $xg=3.94$.

\subsection{Wilson lines and gauge links}
\label{sec:gaugefields}
Once the charge has been generated, we know that the hard and the soft sectors are coupled via the Yang-Mills equations \cite{Yang:1954ek}
\begin{equation}
    D_\mu F^{\mu\nu}=J^\nu,
    \label{8.1}
\end{equation}
where $D_\mu=\partial_\mu-ig[A_\mu,\hspace{3pt} \cdot\hspace{3pt}]$ is the covariant derivative, $F^{\mu\nu}=\partial^\mu A^\nu-\partial^\nu A^\mu-ig[A^\mu,A^\nu]$ is the field strength tensor and $J^\mu$ the color current. Using the light cone coordinates, the conservation of current implies $A^-=0$. Moreover, by working in the covariant gauge we can also impose $A^i=0$, where $i=x,y$. The only component left is therefore $A^+\equiv \alpha$, which can be seen to obey the following Poisson equation
\begin{equation}
    \Delta_\perp \alpha(\mathbf{x}_\perp)=-\rho(\mathbf{x}_\perp).
    \label{9}
\end{equation}
In order to solve this equation, we Fourier-transform both sides of \eqref{9} in order to get
\begin{equation}
    \Tilde{k}^2_\perp \tilde{\alpha}_{n,k}^a=\tilde{\rho}_{n,k}^a.
    \label{9.1}
\end{equation}
The Fourier transformations are easily performed using the 
Fast Fourier Transform (FFT) package in Julia.
We get
\begin{equation}
\tilde{\alpha}_{n,k}^a=\frac{\tilde{\rho}_{n,k}^a}{\Tilde{k}_\perp^2+m^2}
    \label{9.4}
\end{equation}
where the discretized momentum $\Tilde{k}_\perp$ is given by
\begin{equation}
\Tilde{k}_\perp^2=\sum_{i=x,y}\left(\frac{2}{a_\perp}\right)^2\sin^2\left(\frac{k_ia_\perp}{2}\right),
    \label{9.2}
\end{equation}
and we introduced an infrared regulator $m=0.2$ GeV $=1$ fm$^{-1}$  acting as a screening mass. By then, Fourier-transforming back we get  $\alpha(\mathbf{x}_\perp)$ in coordinate space.

Once we obtained a solution for the gauge field in the covariant gauge, the Wilson line for each nucleus is evaluated as
\begin{equation}
    V_\mathbf{x}=\prod_{n=1}^{N_s}\exp\{ig\alpha_{n,\mathbf{x}}^a t^a\}
    \label{9.5}
\end{equation}
and then obtain the gauge links, $U$, for each of the two nuclei as
\begin{equation}
U_{\mathbf{x},i}^{A,B}=V_\mathbf{x}^{A,B}V_{\mathbf{x}+\hat{i}}^{\dagger,A,B}.
\label{9.6}
\end{equation}
In the above expressions $t^a$ are the SU(3) group generators, i.e. the eight Gell-Mann matrices divided by 2. If the point $\mathbf{x}_\perp$ belongs to the edge of the simulation lattice, periodic boundary conditions have been implemented.

\subsection{Initialization of the boost-invariant fields}
At this point, we would like to derive the gauge link for the combined system of the two nuclei immediately after the collision. The gauge field will be the sum of each nucleus' contribution, but since QCD is a non-Abelian gauge theory, the resulting gauge link will not be the product of the gauge links of each nucleus. The total gauge link, $U_{\bm x,i}$, at the position
$\bm x$ in the transverse plane
is determined by solving a  set of eight equations, which are
\begin{equation}
    \text{Tr}[t_a(U_{\mathbf{x},i}^A+U_{\mathbf{x},i}^B)(\mathbb{I}+U_{\mathbf{x},i})-\text{h.c.}]=0,
    \label{10}
\end{equation}
with $a=1,\dots,N_c^2 - 1$.
In the case of  $N_c=2$ there is an exact solution of \eqref{10} for $U_{\mathbf{x},i}$ \cite{Krasnitz:1998ns}. 
Instead, for SU(3)   there is not exact solution and we have to solve \eqref{10} via an iterative method. Using this procedure we calculate the gauge links along the $x$ and $y$ direction, whereas for the longitudinal components we initialize $U_\eta=\mathbb{I}$.
We can also evaluate the color-electric fields, i.e. the canonical momenta with respect to the gauge links, at the initial time. The initial conditions on the lattice are implemented as~\cite{Fukushima:2011nq}
\begin{widetext}
\begin{align}
    &E_x=E_y=0,\nonumber\\
   &E^\eta=-\frac{i}{4ga_\perp^2}\sum_{i=x,y}\left[(U_i(\mathbf{x}_\perp)-\mathbb{I})(U_i^{B,\dagger}(\mathbf{x}_\perp)-U_i^{A,\dagger}(\mathbf{x}_\perp))+(U_i^\dagger(\mathbf{x}_\perp-\hat{i})-\mathbb{I})(U_i^{B}(\mathbf{x}_\perp-\hat{i})-U_i^{A}(\mathbf{x}_\perp-\hat{i}))-h.c.\right].
   \label{11.2}
\end{align}
\end{widetext}

\subsection{Time evolution}
We now discuss the time evolution of the fields whose initialization
has been explained in the previous subsection. The electric fields, are expressed in terms of link variables as \cite{Fukushima:2011nq}
\begin{align}
    \partial_\tau U_i(\mathbf{x})&=\frac{-iga_\perp}{\tau}E^i(\mathbf{x})U_i(\mathbf{x}),\label{eq:parisihatoltoilginger}\\
    \partial_\tau U_\eta(\mathbf{x})&=-iga_\eta\tau E^\eta(\mathbf{x}) U_\eta(\mathbf{x}).
    \label{11.3}
\end{align}
The term $a_\eta$, which denotes the discretization step in the $\eta$-direction.
In order to reduce the discretization error in time, we drive the evolution through a  leapfrog algorithm, i.e. by letting the gauge links and the electric fields evolve in different steps alternatively.
The time evolution for $U_i$ and $U_\eta$ is given by \cite{Fukushima:2011nq}:
\begin{align}
U_i(\tau'')&=\exp\left[-2\Delta \tau\cdot iga_\perp E^i(\tau')/\tau'\right]U_i(\tau),\label{11.4bisAAA}\\
U_\eta(\tau'')&=\exp\left[-2\Delta \tau\cdot ig a_\eta \tau' E^\eta(\tau')\right]U_\eta(\tau),
\label{11.4}
\end{align}
where $\tau'=\tau+\Delta \tau/2$ and $\tau''=\tau+\Delta \tau$. Notice that the exponentiation of the electric field is important, in order to keep the up-to-date gauge links as unitary matrices. In the same fashion, the equations of motion of the electric field are discretized as:
\begin{widetext}
\begin{align}
E^i(\tau')=&E^i(\tau-\Delta \tau)+2\Delta \tau \frac{i}{2ga_\eta^2 a_\perp\tau}[U_{\eta i}(\tau)+U_{-\eta i}(\tau)-(h.c.)]+2\Delta \tau \frac{i\tau}{2g a_\perp^3}\sum_{j\neq i}[U_{ji}(\tau)+U_{-ji}(\tau)-(h.c.)],\nonumber\\
E^\eta(\tau')=&E^\eta(\tau-\Delta \tau)+2\Delta \tau \frac{i}{2g a_\eta a_\perp^2 \tau} \sum_{j=x,y}[U_{j\eta}(\tau)+U_{-j\eta}(\tau)-(h.c.)],\label{12}
 \end{align}
\begin{figure}[t]
\includegraphics[width=\textwidth]{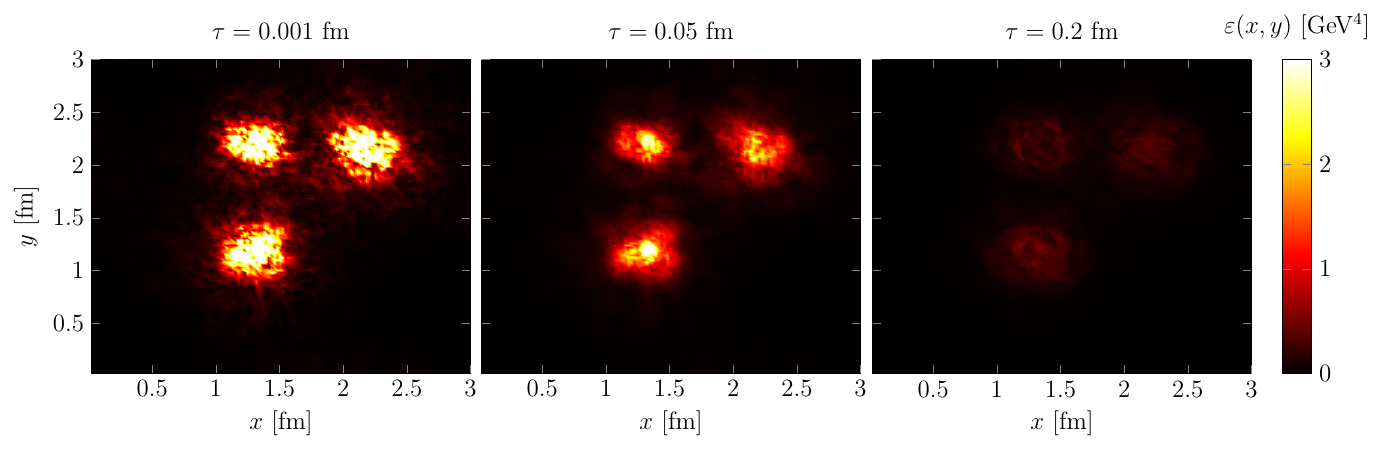}
\caption{Energy density (legend on the right side) produced in a pA collision
in one event, represented in the transverse plane for three values of $\tau$: left panel corresponds to $\tau=0.001$ fm, center panel to $\tau=0.05$ fm, right panel to $\tau=0.2$ fm.}
        \label{Fig:1}
\end{figure}
\end{widetext}
where
\begin{equation}
    U_{\mu\nu}(\mathbf{x})\equiv U_\mu(\mathbf{x})U_\nu (\mathbf{x}+\hat{\mu})U_\mu^\dagger(\mathbf{x}+\hat{\nu})U_\nu^\dagger(\mathbf{x})
    \label{13}
\end{equation}
are the plaquette variables. Notice that in this way the gauge links are initialized at time $\tau=0$, and evolve with integer steps: $\tau=\Delta \tau, 2\Delta \tau, \dots$ Instead, the electric field is initialized at $\tau=\Delta \tau/2$ and proceeds as $\tau=3\Delta \tau/2, 5 \Delta \tau/2,\dots$
The electric fields and gauge fields which we have dealt with so far fit into the definition of the longitudinal/transverse electric and magnetic components of the energy density. In particular, at time $\tau$ we have the electric components given by
\begin{align}
E_L^2(\tau)&\equiv E^{\eta}(\tau)\cdot E^{\eta}(\tau),\nonumber\\
E_T^2(\tau)&\equiv \frac{1}{\tau^2}[E^{x}(\tau)\cdot E^{x}(\tau)+E^{y}(\tau)\cdot E^{y}(\tau)].\label{13.1}
\end{align}
whereas the magnetic components are given by \cite{Fukushima:2011nq}
\begin{align}
    B_L^2(\tau)&=\frac{2}{g^2a_\perp^4}\text{Tr}[\mathbb{I}-U_{xy}(\tau)],\nonumber\\
    B_T^2(\tau)&=\frac{2}{(ga_\eta a_\perp\tau)^2}\sum_{i=x,y}\text{Tr}[\mathbb{I}-U_{\eta i}(\tau)].\label{13.2}
\end{align}
The relations \eqref{13.2} follow easily if one notices that $U_{\mu\nu}\simeq \exp\{-iga_\mu a_\nu F_{\mu\nu}\}$, where $a_\mu$ and $a_\nu$ denote the lattice spacings in the
directions $\mu$ and $\nu$ (no summation over $\mu,\nu$ is intended in the exponent).

The above quantities fit into the definition of $\varepsilon$ \cite{Fukushima:2011nq}
\begin{equation}
    \varepsilon=\langle \text{Tr}[E_L^2+B_L^2+E_T^2+B_T^2]\rangle.
    \label{13.3}
\end{equation}
The angular brackets $\langle \rangle$ denote ensemble average, numerically obtained by averaging over the events.


In Figure \ref{Fig:1} we plot the energy density for a pA collision at three different times. In the left panel, which represents the energy density right after the collision, 
we notice the energy density concentrated over the three lumps 
standing for the three constituent quarks in the proton. As time passes, we notice that the expansion in the transverse plane is negligible,
while the energy density is diluted due to the
longitudinal expansion of the system.

In the lower panel, 
we plot $\varepsilon$ for $\tau=0.15$ fm.
We notice that the 
expansion in the transverse plane is negligible,
while the energy density is diluted due to the
longitudinal expansion of the system.

For completeness,
in Fig.~\ref{Fig:2} we plot $\varepsilon$ versus $\tau$ 
for both pA and AA collisions,
averaged over $100$ events and over the transverse plane. In order to perform the transverse plane average consistently in the pA case, we have used the thickness function of the proton $T_p(\mathbf{x}_\perp)$ as a weight, i.e. we plot $\langle \langle \text{Tr}[E_L^2+B_L^2+E_T^2+B_T^2] \rangle \rangle$ where $\langle \langle A \rangle \rangle$ for any (transverse--space dependent) quantity $A$ is defined as 
\begin{equation}
\langle\langle
 A
\rangle\rangle =
\int d^2\mathbf{x}_\perp\langle A(\mathbf{x}_\perp)\;T_p(\mathbf{x}_\perp)\rangle.
\label{eq:rai3_22_02}
\end{equation} 
In the above definition the function $T_p(\mathbf{x}_\perp)$, defined in Eq.~\eqref{eq:bd1}, takes into account the initial distribution of the HQs in the configuration space. Moreover, the average in the right hand side of \eqref{eq:rai3_22_02} is over the numerical events.

\begin{figure}[ht]
    \centering
\includegraphics[width=0.9\linewidth]{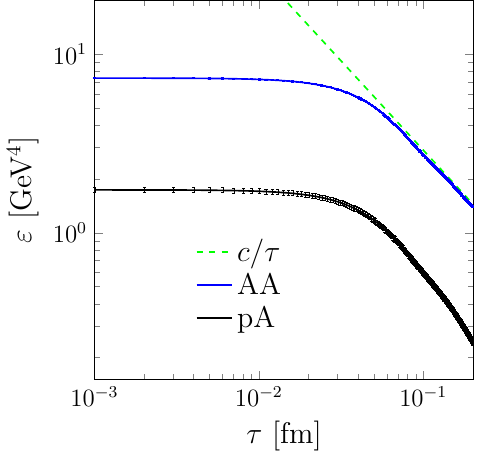}
    \caption{Energy density $\varepsilon$, versus 
    proper time, averaged over the transverse plane (for the pA case, the average over the transverse plane has been performed using $T_p(\mathbf{x}_\perp)$ as a weight, see main text). Error bars are also shown (for the AA case the error is smaller than the line width).}
    \label{Fig:2}
\end{figure}

The results have been obtained for a square grid of transverse size
$L=2$ fm, with $32\times 32$ lattice points. Along with the event averages, here and for all the plots that will follow, also plot the statistical uncertainty as an error band, obtained as the standard deviation of the mean value: $\sigma_{\text{avg}}=\sigma/\sqrt{N_{\text{events}}}$.
We notice that the free-streaming regime, $\varepsilon\sim 1/\tau$, sets in at $\tau\approx 0.1$ fm. For reference, the initial values of $\varepsilon$ are, expressed in terms of energy over volume, around 240 GeV/fm$^3$ for pA and 960 GeV/fm$^3$ for AA.

\section{Quarks of infinite mass in glasma}
\label{Quarks of infinite mass in Glasma}
Next we turn to 
the diffusion of HQs in the evolving glasma fields.
This problem has attracted a lot of interest recently,
see for example \cite{Ruggieri:2018ies,Ruggieri:2018rzi,Sun:2019fud,Liu:2019lac,Jamal:2020fxo,Liu:2020cpj,Sun:2021req,Khowal:2021zoo, Pooja:2022ojj,Pooja:2024rnn, Pooja:2024rsw, Avramescu:2024poa, Avramescu:2024xts} as well as references
therein.
So far, this problem has been solved using a classical approximation 
in which
the equations of motion for the HQs are the Wong equations~ \cite{Wong:1970fu, Heinz:1984yq}.
In this work, we limit ourselves to analyze the diffusion of heavy quarks
in the very-large mass limit, similarly to
what has been done in \cite{Boguslavski:2020tqz}. 

\subsection{Momentum broadening}
One of the quantities we want to study 
is the momentum broadening, $\delta p_i^2$, 
of the HQs. It has already been studied, for instance, in~\cite{Ruggieri:2018ies,Ruggieri:2018rzi,Sun:2019fud,Liu:2019lac,Jamal:2020fxo,Liu:2020cpj,Sun:2021req,Khowal:2021zoo, Pooja:2022ojj,Pooja:2024rnn, Pooja:2024rsw, Avramescu:2024poa, Avramescu:2024xts}.
We analyze this quantity in the present study, bringing the novelty of introducing longitudinal fluctuations. As a benchmark, we firstly show the results obtained without fluctuations.

This quantity is defined as
\begin{equation}
\delta p_i^2(\tau)\equiv p_i^2(\tau)-p_i^2(\tau_{\text{form}}),~~~
i=x,y,z,
\label{eq14.1}
\end{equation}
where $\tau_{\text{form}}$ is the formation time of the HQ and $p^i(\tau_{\text{form}})$ the $i$-th component of the momentum at the formation time. 
In order to compute the momentum broadening, 
we need to specify the initial distribution of the HQs in the
configuration space. To this end, 
we assume that
HQs are produced near the hotspots of the energy density (see Fig.~\ref{Fig:1}). The idea behind this claim is that heavy quarks can only be produced in the regions with highest energy density. This is justified since HQs are produced by hard scatterings, them being quite massive, so their formation rate must be higher in correspondence to a large number of binary collisions within the two colliding nuclei (here p and A).
Hence, it is reasonable to distribute the
HQs according to the $T_p(\mathbf{x}_\perp)$ in Eq.~\eqref{eq:bd1}.
Following the same arguments presented
in~\cite{Avramescu:2023qvv}, from the Wong equations we can write the
momentum broadening of each heavy quark,
in the $M\rightarrow\infty$ limit,
as
\begin{widetext}
\begin{eqnarray}
\delta p_L^2(\tau) &=&
g^2\int_{0}^\tau d\tau' \int_{0}^\tau d\tau'' \text{Tr}[E_z(\tau')E_z(\tau'')],
\label{eq:filmnoiososuamazonprime}\\
\delta p_T^2(\tau) &=&
g^2\int_{0}^\tau d\tau' \int_{0}^\tau d\tau'' 
\frac{1}{\tau^\prime \tau^{\prime\prime}}
\text{Tr}[E_x(\tau')E_x(\tau'')+E_y(\tau')E_y(\tau'')],
\label{eq:italia1_21_34}
\end{eqnarray}
where we took the formation time of the infinitely massive HQs
equal to zero, since $\tau_{\text{form}}=O(1/M)$ and $M\to \infty$.
The integrands in the right-hand side of 
Eqs.~\eqref{eq:filmnoiososuamazonprime}
and~\eqref{eq:italia1_21_34} depend also on the transverse plane coordinates.
Integrating over the whole transverse plane and ensemble-averaging 
Eqs.~\eqref{eq:filmnoiososuamazonprime} and~\eqref{eq:italia1_21_34},
results in
\begin{eqnarray}
\langle\langle\delta p_L^2(\tau)\rangle\rangle &=&
g^2\int_{0}^\tau d\tau' \int_{0}^\tau d\tau'' \int d^2\mathbf{x}_\perp
\langle
\text{Tr}[E_z(\tau')E_z(\tau'')] T_p(\mathbf{x}_\perp)\rangle,
\label{eq:oraguardiquesto}\\
\langle\langle\delta p_T^2(\tau)\rangle\rangle &=&
g^2\int_{0}^\tau d\tau' \int_{0}^\tau d\tau'' \int d^2\mathbf{x}_\perp
\frac{1}{\tau^\prime \tau^{\prime\prime}}
\langle \text{Tr}[E_x(\tau')E_x(\tau'')+E_y(\tau')E_y(\tau'')]T_p(\mathbf{x}_\perp)\rangle.
\label{eq15}
\end{eqnarray}
In writing Eqs.~\eqref{eq:oraguardiquesto} and~\eqref{eq15}
we took advantage of the fact that the HQs are infinitely massive,
therefore their distribution in the configuration space does not 
change in time.

\end{widetext}
The equations~\eqref{eq:oraguardiquesto}
and~\eqref{eq15} are formally similar to what we would obtain
in the case of a standard Langevin equation
for a purely diffusive motion: $dp^i/dt=\xi^i$,
where $\xi^i$ denotes the random force. In fact, in this case
we would obtain~\cite{Liu:2020cpj}
\begin{equation}
\langle \delta p_i^2\rangle = \int_0^t dt_1\int_0^t dt_2
\langle
\xi_i(t_1)\xi_i(t_2)
\rangle.
\label{eq:tuttivoglionoFrattesi}
\end{equation}
We can indeed notice the similarity among \eqref{eq:tuttivoglionoFrattesi}
and Eqs.~\eqref{eq:oraguardiquesto} and~\eqref{eq15} after realizing that
the role of the random force in the context of HQs
is taken by the force exerted by the color-electric field. Within a Langevin dynamics, one can in principle include ‘‘memory effects'', which here are not considered, in order to account for coherent field-induced dynamics \cite{Ruggieri:2022kxv}. Moreover, one can also include radiation effects, but these have been shown to be basically irrelevant within the dynamics of heavy quarks in glasma \cite{Liu:2020cpj}.
We also notice that in the infinite mass limit
Eqs.~\eqref{eq:oraguardiquesto} and~\eqref{eq15} are exact,
in the sense that the color-magnetic field gives no contribution
in this limit as its effect is proportional to the
velocity of the quarks, which vanishes for $M\to \infty$.
For $i=x,y$ the resulting equation for the momenta shift can be evaluated directly from the $E_x$ and $E_y$ fields. On the other hand, 
in principle we need to 
explicitly compute the $E_z$ component of the color-electric field
from the $(\tau,\eta)$ components. 
This can be easily achieved 
as 
under the change of coordinates 
from $x^{\mu'}=(\tau,\eta)$ to $x^\mu=(t,z)$ in Eq.~\eqref{15.2} we have
\begin{equation}
    E_z=F_{0z}=\frac{\partial x^{\mu '}}{\partial t}\frac{\partial x^{\nu'}}{\partial z}F_{\mu' \nu'}=F_{\tau \eta}=E_\eta.
    \label{15.1}
\end{equation}
Consequently,
the knowledge of the $\eta-$component of the field, which we 
easily extract from the
Yang-Mills equations, is enough to compute the momentum spreading
along the $z$ direction.

\begin{figure}[t!]
    \centering
    \includegraphics[width=\linewidth]{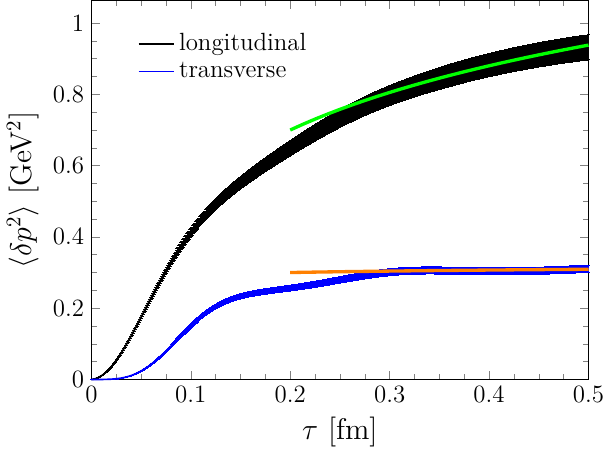}
    \caption{Longitudinal momentum broadening $\langle \delta p_L^2(\tau)\rangle$ (black), and transverse momentum broadening
$\langle \delta p_T^2(\tau)\rangle$ (blue) for a pA collision, along with the corresponding error bars. The green line
corresponds to the function $a+b\log(\tau/\bar{\tau})$ where we put
$\bar{\tau}=0.4$ fm, $a=0.88$ GeV$^2$ and $b=0.26$ GeV$^2$.
Similarly, the orange solid line represents the function
$c + d \log(\tau/\tilde{\tau})$ where we put
$\tilde{\tau}=0.2$ fm, $c=0.30$ GeV$^2$ and 
$d=0.01$ GeV$^2$.}
    \label{Fig:3}
\end{figure}

In Fig.~\ref{Fig:3} we show $\langle \delta p_L^2(\tau)\rangle$ and
$\langle \delta p_T^2(\tau)\rangle$ versus proper time, calculations
correspond to an average over $100$ events, for a transverse plane with $32\times 32$ points and transverse size $L=2$ fm.
The evolution is shown up to $\tau=0.5$ fm. We notice that the transverse momentum shift is substantially lower than the longitudinal one: this is a consequence of the predominance of the longitudinal color-electric fields in the first instants after the collision. The anisotropy of the evolving glasma,
which reflects onto the difference between the longitudinal and the
transverse field correlators, is transmitted to the momentum broadenings. We find that, as a result of the diffusion of the HQs in the evolving glasma fields, both $\langle \delta p_L^2(\tau)\rangle$ and $\langle \delta p_T^2(\tau)\rangle$ grow up with proper time; after a short transient lasting up to $\tau\approx 0.3$ fm,
the growth of $\langle \delta p_T^2(\tau)\rangle$ drastically slows down,
while that of $\langle \delta p_L^2(\tau)\rangle$ remains substantial
for the whole time range considered here. The values obtained can be compared, for example, with those found in \cite{Avramescu:2023qvv} for AA collisions: our momenta shifts are smaller since we are dealing with pA collisions, in which the charge density (and hence the energy density) is smaller.

The late-time behavior of $\langle\delta p_L^2(\tau)\rangle$
and $\langle\delta p_T^2(\tau)\rangle$
can be interpreted by assuming a $\delta-$like correlator
of the color-electric fields. For instance, let us assume that for $E_z$ we have
\begin{equation}
\langle E_z(\tau^\prime)E_z(\tau^{\prime \prime})\rangle = 
\frac{  \langle E_z^2(\tau^\prime)\rangle}{Q_s}
\delta(\tau^\prime - \tau^{\prime\prime}),
\label{eq:sonobravisullavoro}
\end{equation}
where the overall $1/Q_s$ 
on the right hand side of the above equation
is added in order to balance the units of the 
$\delta$ function. The particular choice of $Q_s$
in the equation is due to the fact that $Q_s$ is the only
energy scale in the MV model.
Using the ansatz~\eqref{eq:sonobravisullavoro} in Eq.~\eqref{eq:oraguardiquesto}
we easily get
\begin{equation}
\langle \delta p_L^2(\tau)\rangle_{M\to \infty} \approx g^2 Q_s^{-1}
\int_{\bar{\tau}}^\tau d\tau^\prime \langle E_z^2(\tau^\prime)\rangle,
\label{eq:messicaniinfuga}
\end{equation}
where we cut the time-integral down to $\bar{\tau}$, which represents a value of the proper time of the
order of $1/Q_s$. We introduce it since 
we are interested in the late-time behavior only, indeed it is
large enough that  
allows us to use $\langle E_z^2(\tau^\prime)\rangle  \propto Q_s^4/(Q_s\tau^\prime)$,
in agreement with the late-time behavior of the energy
density, see Fig.~\ref{Fig:2}.
We have also inserted
the needed powers of $Q_s$ in order to get the correct dimension of energy to the fourth power.\footnote{For this argument we do not need to be precise and fix
the overall coefficient, as we are only interested in
extracting the late-time  behavior of the momentum
broadening.}
We thus get
\begin{equation}
\langle \delta p_L^2(\tau)\rangle_{M\to \infty} 
\propto
g^2 Q_s^2
\log(\tau/\bar{\tau}),~~~\text{for }\tau/\bar{\tau} >1.
\label{eq:conoscebeneilmestiere}
\end{equation}
The late time behavior~\eqref{eq:conoscebeneilmestiere} of 
$\langle \delta p_L^2(\tau)\rangle$ 
(and similar calculations hold for $\langle \delta p_T^2(\tau)\rangle$)
is in fair agreement with the results 
of the full numerical simulations,
see orange and green lines in Fig.~\ref{Fig:3}. Here, by ‘‘late time'' we refer to timescales which are at the upper limit of the validity of the glasma framework, i.e. around $\tau\sim 0.4$ fm.
As a final comment on the results shown in
Fig.~\ref{Fig:3}, we notice that we measure some tiny fluctuations
of both $\langle \delta p_L^2(\tau)\rangle$ and 
$\langle \delta p_T^2(\tau)\rangle$, similarly to what observed  in~\cite{Avramescu:2023qvv}. The amplitude of these fluctuations
is however very small in comparison with
the bulk value of these quantities, thus they are practically irrelevant.

\subsection{Angular momentum broadening}
The momentum anisotropy is also transmitted to other
observables of the HQs.
In particular,
one can look at the angular momentum anisotropy parameter, 
$\Delta_2$, defined as~\cite{Pooja:2022ojj}
\begin{equation}
    \Delta_2\equiv \frac{\langle L_x^2-L_z^2\rangle}{\langle L_x^2+L_z^2\rangle},
    \label{22}
\end{equation}
where $L_i$, with $i=x,y,z$, denotes the $i^{th}$ component of the
orbital angular momentum of the HQ. $\Delta_2$ therefore measures the anisotropy
of the fluctuations of the orbital angular momentum of the HQs (the fluctuations of the HQ spin are suppressed by a power
of $M$~\cite{Pooja:2022ojj}). The notation makes clear that this quantity is basically a second moment of the HQ angular momentum distribution. It is also worth noting that $\Delta_2$ receives contribution from both the anisotropy of the color
fields and from the geometry of the system~\cite{Pooja:2022ojj}. 

Such quantity can be rewritten 
in terms of the averages of the squared components of the momentum of the HQs. This can be done following the same steps given in~\cite{Pooja:2022ojj} and assuming that momenta and positions of the HQs are not correlated with each other: this holds since we are working in the static limit for HQs. We have
\begin{align}
\langle L_x^2\rangle&=\langle (yp_z-zp_y)^2\rangle\simeq \langle y^2\rangle\langle p_z^2\rangle+\langle z^2\rangle\langle p_y^2\rangle,\nonumber\\
\langle L_z^2\rangle&=\langle (xp_y-yp_x)^2\rangle\simeq \langle x^2\rangle\langle p_y^2\rangle+\langle y^2\rangle \langle p_x^2\rangle.
\label{38}
\end{align}
The above expressions allow us to write
\begin{equation}
    \Delta_2=\frac{[\langle z^2\rangle-\langle x^2\rangle-\langle y^2\rangle]\langle p_T^2\rangle/2+\langle y^2\rangle \langle p_z^2\rangle}{[\langle z^2\rangle+\langle x^2\rangle+\langle y^2\rangle]\langle p_T^2\rangle/2+\langle y^2\rangle \langle p_z^2\rangle}.
    \label{39}
\end{equation}

This expression can be further specialized
to the case of a pA collision as follows.
Firstly, 
we estimate the mean squared value of the longitudinal $\langle z^2\rangle$ and of the transverse coordinates $\langle x^2\rangle$ and $\langle y^2\rangle$. 
As far as the extension along the beam axis $z$ is concerned, we will assume that the HQs are generated uniformly along the whole $\eta$ interval. This means that $\langle z^2\rangle$ will be given by:
\begin{equation}
    \langle z^2\rangle =\frac{\displaystyle{\int_{-L_\eta/2}^{L_\eta/2}}d\eta\, (\tau \sinh \eta)^2}{\displaystyle{\int_{-L_\eta/2}^{L_\eta/2}}d\eta}=\frac12 \tau^2 \left(\frac{\sinh L_\eta}{L_\eta}-1\right).
    \label{40}
\end{equation}
We fix the $\eta$ extension as $L_\eta=2$, corresponding to the rapidity range $\eta \in[-1,1]$. For what concerns the transverse coordinates, firstly
we assume that the geometrical distribution of the HQs in the transverse
plane follows the same profile of the initial energy density,
which in turn resembles that of the color charges of the colliding proton.
This is a fair approximation considering that most HQs will be produced
by hard QCD scatterings where the density of the color charges is larger,
as well as the fact that the transverse expansion is not so important
in the pre-equilibrium stage, see for example Fig.~\ref{Fig:1}.
Within these reasonable approximations, we can estimate 
$\langle \mathbf{x}_\perp^2 \rangle =\langle x^2 + y^2 \rangle $. Firstly, we fix the location of the
three constituent quarks, $\mathbf{x}_i$, 
and we average over the transverse coordinates using the thickness function $T_p( \mathbf{x}_\perp)$ in \eqref{eq:bd1}, getting
\begin{equation}
    \langle \mathbf{x}_\perp^2\rangle_{\mathbf{x}_i}=\frac{\displaystyle{\int} d^2\mathbf{x}_\perp\, \mathbf{x}_\perp^2\, T_p(\mathbf{x}_\perp)}{\displaystyle{\int} d^2\mathbf{x}_\perp\, T_p(\mathbf{x}_\perp)}=2B_q+\frac13 \sum_{i=1}^3 \mathbf{x}_i^2.
    \label{41}
\end{equation}
Then 
an average over the locations of the constituent quarks,
$\mathbf{x}_i$, is needed: we perform this
using the distribution $T_{cq}(\mathbf{x}_i)$ in~\eqref{eq:bd2}, resulting in 
\begin{equation}
    \langle \mathbf{x}_\perp^2\rangle=\frac{\displaystyle{\int} d^2\mathbf{x}_i\, \langle \mathbf{x}_\perp^2\rangle_{\mathbf{x}_i} T_{cq}(\mathbf{x}_i)}{\displaystyle{\int} d^2\mathbf{x}_i\, T_{cq}(\mathbf{x}_i)} =2B_q+2B_{cq}.
    \label{42}
\end{equation}
Hence,
\begin{equation}
    \langle x^2\rangle= \langle y^2\rangle=\frac12 \langle \mathbf{x}_\perp^2\rangle=B_q+B_{cq}.
    \label{43}
\end{equation}
Using the results~\eqref{40} and~\eqref{43} 
in Eq.~\eqref{39}, along with the momenta shifts we previously got from \eqref{eq:oraguardiquesto} and \eqref{eq15},
we can compute $\Delta_2$ for a pA collision.

We have already mentioned that a nonzero $\Delta_2$ may not only be due to the anisotropy of the fields (which is reflected on to a anisotropic momentum distribution),
but entails a geometric contribution as well. The latter is related to the shape of the fireball and could be nonzero also in case momenta are isotropic. 
In order to 
extract the geometric contribution to $\Delta_2$, we start with Eq.~\eqref{39} in which we evaluate $\langle p_z^2\rangle$ as follows. Since we have
\begin{equation}
p_z=p_T\sinh y\simeq p_T \sinh \eta,
    \label{eq:pz_to_pTsineta}
\end{equation}
by assuming once again that the HQs are uniformly spread in $\eta$ we have:

\begin{equation}
    \langle p_z^2\rangle =\frac{\displaystyle{\int_{-L_\eta/2}^{L_\eta/2}}d\eta\, \langle p_T^2\rangle (\sinh \eta)^2}{\displaystyle{\int_{-L_\eta/2}^{L_\eta/2}}d\eta}=\frac12 \langle p_T^2\rangle \left(\frac{\sinh L_\eta}{L_\eta}-1\right).
    \label{eq:avgd_pz_to_pTsineta}
\end{equation}
This allows us to define
the geometric component of $\Delta_2$, we call it $\Delta_2^{\text{geom}}$, namely 
\begin{align}
&\Delta_2^{\text{geom}}=\frac{\langle z^2\rangle+\langle x^2\rangle-\langle y^2\rangle+\langle y^2\rangle\left(\sinh L_\eta/L_\eta-1\right)}{\langle z^2\rangle+\langle x^2\rangle+\langle y^2\rangle+\langle y^2\rangle\left(\sinh L_\eta/L_\eta-1\right)}=\nonumber\\
&=\frac{[\frac12 \tau^2+(B_q+B_{cq})]\left(\sinh L_\eta/L_\eta-1\right)}{[\frac12 \tau^2+(B_q+B_{cq})]\left(\sinh L_\eta/L_\eta-1\right)+2(B_q+B_{cq})}.
    \label{44}
\end{align}

In Figure~\ref{Fig:4} we show our results for both $\Delta_2$ and $\Delta_2-\Delta_2^{\text{geom}}$. 
The calculations correspond to a pA collision for a $32\times 32$ lattice
and an average over $100$ events.
From Eq.~\eqref{44} we get $\Delta_2^{\text{geom}}(\tau=0)=
(\sinh L_\eta/L_\eta-1)/(\sinh L_\eta/L_\eta+1)\simeq 0.289$, therefore we have a geometric correction already at initial time. Notice also that $\Delta_2^{\text{geom}}(\tau=0)$ is independent with respect to the widths of the proton charge, namely $B_q$ and $B_{cq}$. For times up to $\tau \sim 0.15 $ fm, $\Delta_2^{\text{geom}}(\tau)$ varies with time but quite slowly (see the $\tau^2$ term in \eqref{44}), hence $\Delta_2^{\text{geom}}(\tau)\simeq \Delta_2^{\text{geom}}(\tau=0)$ and the two curves in Figure~\ref{Fig:4} differ basically by a constant. During this time we observe most of the damping for both curves.
For even larger times we see that $\Delta_2$ (black curve) shows an increasing behavior, but a look at $\Delta_2-\Delta_2^{\text{geom}}$ in the same time scales shows that the increase in angular momentum anisotropy is likely due to an expansion of the system, mostly along the longitudinal direction. Indeed, by subtracting the geometric component we see that the pure anisotropy of the glasma fields (blue curve) remains constant and small over time.


\begin{figure}
    \centering
    \includegraphics[width=\linewidth]{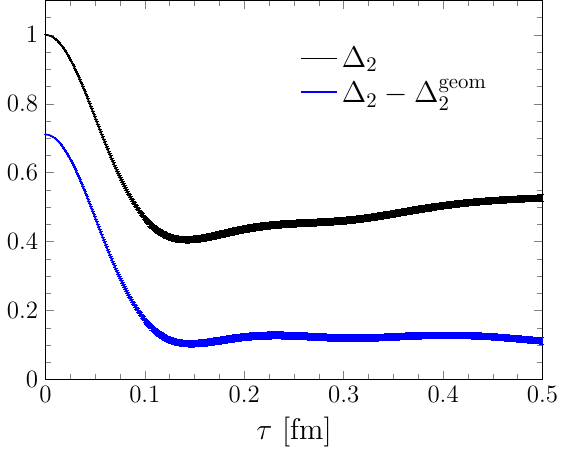}
    \caption{$\Delta_2$ (black) and  $\Delta_2-\Delta_2^{\text{geom}}$ (blue) versus proper time
    for pA collisions, along with the corresponding error bars.}
    \label{Fig:4}
\end{figure}

\section{Inclusion of field fluctuations}
\label{Initial state fluctuations}

As already explained in the Introduction,
in relativistic nuclear collisions 
one clearly does not have exact boost-invariance. 
In particular, 
the glasma is the solution of the Yang-Mills equations
only in the case of weak coupling $g$: 
for realistic values of $g$ it is likely that quantum fluctuations develop, with a strength that 
increases with $g$.  
For instance,
in \cite{Epelbaum:2013ekf} it has been shown that increasing $g$ leads to a larger contribution
of fluctuations to the longitudinal pressures. 
Since a finite value of $g$ is relevant for realistic collisions,
it is meaningful to repeat the analysis presented in the previous section to the case in which we add fluctuations on top of the glasma.
We construct the fluctuations by adding terms to the boost-invariant electric fields at initial time. These additions
satisfy the Gauss' law
\begin{equation}
    D_i E_i+D_\eta E_\eta=0.
    \label{45}
\end{equation}
Firstly, we evaluate random configurations $\xi_i(\mathbf{x}_\perp)$ ($i=x,y$) such that
\begin{equation}
    \langle \xi_i(\mathbf{x}_\perp) \xi_j(\mathbf{y}_\perp)\rangle=\delta_{ij} \delta^{(2)}(\mathbf{x}_\perp-\mathbf{y}_\perp),
    \label{46}
\end{equation}
which are operatively generated by sampling random Gaussian numbers with zero average and standard deviation $1/a_\perp$. 
After that, the additional terms to the 
glasma color-electric fields are evaluated as \cite{Fukushima:2011nq,Romatschke:2006nk}
\begin{align}
        \delta& E^i(\mathbf{x}_\perp,\eta)=-\partial_\eta F(\eta)\xi_i(\mathbf{x}_\perp),\label{eq:ottimathuram}\\
        \delta& E^\eta(\mathbf{x}_\perp,\eta)=F(\eta)\sum_{i=x,y}D_i \xi_i(\mathbf{x}_\perp).\label{47}
    \end{align}
The information about the $\eta-$distribution of the
fluctuations is encoded in the function $F(\eta)$.
By construction, these fluctuations satisfy the Gauss law
constraint~\eqref{45}. 
The above relations are discretized on the lattice as~\cite{Fukushima:2011nq}
\begin{align}
\delta& E^i(\mathbf{x}_\perp,\eta)=a_\eta^{-1}[F(\eta-a_\eta)-F(\eta)]\xi_i(\mathbf{x}_\perp),\\
\delta& E^\eta(\mathbf{x}_\perp,\eta)=
        -a_\perp^{-1}F(\eta)\sum_{i=x,y}[
        \Xi^i(\mathbf{x}_\perp)
        -\xi^i(\mathbf{x}_\perp)],\label{47bis}
    \end{align}
where
\begin{equation}
\Xi^i(\mathbf{x}_\perp) = 
U_i^\dagger(\mathbf{x}_\perp-\hat{i})\xi_i(\mathbf{x}_\perp-\hat{i})U_i(\mathbf{x}_\perp-\hat{i}).
\end{equation}
In the above equations, $a_\eta=L_\eta/N_\eta$ is the lattice spacing along the $\eta$ direction.
The term  $\Xi^i(\mathbf{x}_\perp)
        -\xi^i(\mathbf{x}_\perp)$
is the covariant derivative of $\xi^i$, expressed in terms of the gauge links. Moreover, also in the longitudinal direction, periodic boundary conditions have been implemented for points at the edge of the $\eta$ domain.
As far as the choice of the rapidity-dependent term $F(\eta)$ is concerned, one may consider different approaches. For instance, in~\cite{Romatschke:2006nk} 
the authors
implemented a model in which 
$F(\eta)$ 
is a Gaussian random number with zero mean and 
standard deviation equal to one.
In this work, we instead follow a
simpler model of fluctuations, similar to the one 
introduced in~\cite{Fukushima:2011nq}, 
in which $F(\eta)$
takes the form
\begin{equation}
F(\eta)=\frac{\Delta}{N_\perp}\sum_{n\in \mathcal{I}}
\frac{1}{|\mathcal{I}|}
 \cos\left(\frac{2\pi n \eta}{L_\eta}\right),
    \label{48_bis}
\end{equation}
where $\mathcal{I} \in \mathbb{N}$ denotes a set of positive
integers with cardinality $|\mathcal{I}|$;
the higher the order of the 
harmonics that contribute to $F(\eta)$ in~\eqref{48_bis}
the higher the energy carried by the 
transverse fields of the
fluctuations, see the $\eta$-derivative in Eq.~\eqref{eq:ottimathuram}. Notice the presence of a free 
parameter, $\Delta$, which encodes the strength of the
fluctuating fields.

\subsection{Momentum broadening fluctuations\label{sec:prescel}}

In this subsection we present our results on the impact of the initial fluctuations on the momentum broadening of the HQs in the pre-equilibrium stage of pA collisions.
The fluctuations are added 
on top of the boost-invariant glasma at 
$\tau\equiv\tau_0=0.05$ fm;
we checked that the results are not significantly
affected by changing $\tau_0$ in the range
$[0.01,0.05]$ fm. The plots which follow have been obtained in a $32\times 32 \times 32$ lattice, as an average over $30$ events, for different values of $\Delta$.

\begin{figure}
\centering
\includegraphics[width=\linewidth]{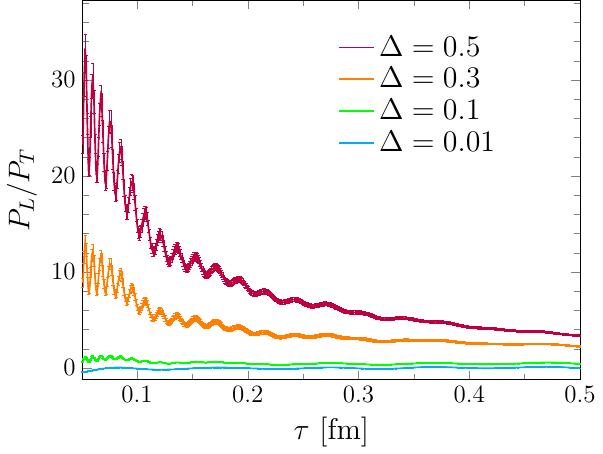}
\caption{The ratio of the longitudinal pressure over the transverse pressure $P_L/P_T$, versus proper time for different values of $\Delta$. Error bars are also shown.}
\label{Fig:PL_over_PT}
\end{figure}

\begin{figure}
\centering
\includegraphics[width=\linewidth]{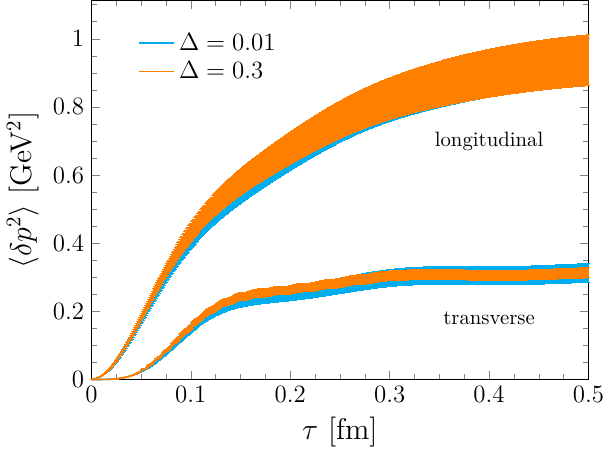}
\includegraphics[width=\linewidth]{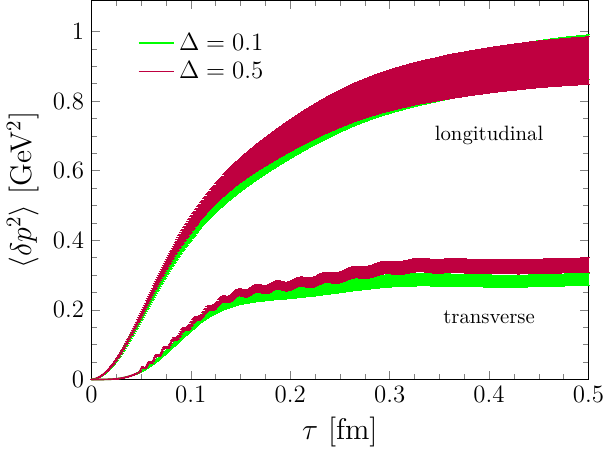}
\caption{The longitudinal and transverse momentum shifts, $\langle \delta p_L^2(\tau)\rangle$ and
$\langle \delta p_T^2(\tau)\rangle$,
versus proper time for different values of $\Delta$. We also show the corresponding error bars.
We used $\mathcal{I}=\{1,\dots,10\}$ in Eq.~\eqref{48_bis}. The four curves have been split into two plots for the sake of clarity.}
\label{Fig:shifts_pt3}
\end{figure}

In Fig.~\ref{Fig:PL_over_PT} we show, for the set of harmonics $\mathcal{I}=\{1,\dots,10\}$ and several
values of $\Delta$, the ratio of longitudinal pressure over transverse pressure, $P_L/P_T$, versus proper time. These quantities are defined as \cite{Fukushima:2011nq}:
\begin{align}
    P_L&=\langle \langle\text{Tr}[E_T^2+B_T^2-E_L^2-B_L^2]\rangle\rangle.\\
    P_T&=\langle\langle \text{Tr}[E_L^2+B_L^2]\rangle\rangle.
\end{align}
where the ensemble average is performed as in \eqref{eq:rai3_22_02}. The main 
purpose of the results shown in Fig.~\ref{Fig:PL_over_PT}
is to highlight the net effect of the fluctuations
on the bulk gluon fields. 
Indeed, for the highest values of $\Delta$, we find a non-trivial behavior of $P_L/P_T$, which 
is quite different from the one we find for smaller values of $\Delta$ as well as for $\Delta=0$ (the result for $\Delta=0$ is not shown, since we checked that it is indistinguishable from the $\Delta=0.01$ result). In particular, for large $\Delta$, $P_L/P_T$ stands significantly higher than the close-to-zero values obtained for small $\Delta$. We remark that the largest value of $\Delta$ used in
Fig.~\ref{Fig:PL_over_PT}, that is $\Delta=0.5$, is probably
too large as it corresponds to have a system whose energy is carried
half by glasma and half by fluctuations. Nevertheless,
it is shown to illustrate that fluctuations are actually introduced
in the bulk and substantially modify it.

Next, we turn to
the momentum broadening. 
In Fig.~\ref{Fig:shifts_pt3} we plot 
$\langle \langle\delta p_L^2(\tau)\rangle\rangle$ and
$\langle \langle\delta p_T^2(\tau)\rangle\rangle$
versus proper time. We show results for the same set of harmonics $\mathcal{I}=\{1,\dots,10\}$ and for the same parameters $\Delta$ we have already considered in Fig. \ref{Fig:PL_over_PT}: we have split the datasets into two different plots, just for the sake of clarity. In addition to the averaging over the transverse plane (as in \eqref{eq:rai3_22_02}),
we average on the space-time rapidity of the HQs,
restricting ourselves to the rapidity range $[-1,1]$.

We find no net effect of the rapidity fluctuations on the momenta shifts, since each result lies within the error bars of all the others, including the highest value $\Delta=0.5$. Our conclusion is that the fluctuations do not affect 
substantially
the HQ momenta shifts in the static limit.
Our interpretation is that momentum broadening
is related, in the static limit, to the time-correlator
of the color-electric fields, see Eqs.~\eqref{eq:oraguardiquesto}
and~\eqref{eq15}, and these correlators are not
very much affected by the longitudinal 
fluctuations, which add a non-trivial dependence of the
fields on $\eta$ and mostly modify correlations along the $\eta$-direction~\cite{Ruggieri:2017ioa}.

\section{Conclusions}

We studied the diffusion of heavy quarks
in the early stage of high-energy proton-nucleus
collisions. Our initial condition is based on 
the glasma picture,
and includes event-by-event
fluctuations of the color charges in the proton, 
as well as fluctuations that break the boost invariance.
We performed $3+1$D real-time statistical simulations 
of the SU(3) Yang-Mills fields produced immediately
after the collision.
Both the modeling of the proton and the
longitudinal fluctuations for an expanding geometry
in SU(3) have not been considered before
in the literature in the context of heavy quarks.
For simplicity, we limited ourselves to study the large-mass
limit of the heavy quarks: in this limit, 
the solution of the Wong equations for the momentum diffusion
amounts to computing the time-correlator of the
color-electric fields.
As a model of initial state field fluctuations,
we adopted a simple superposition of several harmonics~\cite{Fukushima:2011nq}, with a free 
parameter, $\Delta$, that measures the strength of the
fluctuating fields, see Eq.~\eqref{48_bis}.
The fact that the strength of the fluctuating
fields scales as $\Delta/|\mathcal{I}|$ in 
Eq.~\eqref{48_bis} allows us to keep the 
initial energy density carried by the fluctuations
unaffected by the inclusion of more harmonics.

We found that 
momentum broadening
both in longitudinal direction and in the transverse plane, $\langle\delta p_L^2\rangle$ and $\langle\delta p_T^2\rangle$ respectively, are unaffected (within error bars) 
by the inclusion of $\eta$-dependent fluctuations. 
Our interpretation of this result is that the fluctuations
do not alter in a substantial way the time  correlations of the color-electric fields, that are directly related
to momentum broadening.
Hence, our conclusion is
that the initial state fluctuations,
at least in the form introduced within our study,
might be not sufficient
to achieve isotropization 
of heavy quark momenta
within the early stage. This occurs despite the 
fact that for the (illustrative) case of intense fluctuations
considered in our work, the quantity $P_L/P_T$,
which is a measure of the amount of anisotropy
of the bulk gluon fields, may be an order of magnitude larger than one.

Additionally, we computed the anisotropy of the
angular momentum fluctuations, 
that we quantified by the coefficient $\Delta_2$
introduced in~\cite{Pooja:2022ojj}.
In~\cite{Pooja:2022ojj} the $\Delta_2$ has been studied
for a static geometry, within the gauge group SU(2) and
without initial state fluctuations, hence the present work
improves these several aspects
with respect to \cite{Pooja:2022ojj}.
We found that $\Delta_2$ remains considerably large
during the whole early stage: 
in the time range where 
the picture based on
an evolving glasma is phenomenologically
relevant, that is 
for $\tau$ in the
range $[0.2,0.4]$ fm, $\Delta_2$ remains 
in the range $[0.4,0.6]$, signaling a substantial amount
of anisotropic angular momentum fluctuations. For larger times we can grasp a linear trend for $\Delta_2$, however such behavior not only may not be accessible phenomenologically, but it is also likely due to the expansion of the system, rather than a genuine feature of the glasma fields.

The work we presented here can be improved in several ways. Certainly, the most important improvement would be the inclusion of dynamical heavy quarks, hence going beyond the large-mass limit. That should be accompanied by a proper initialization of the heavy quarks in momentum space, as already done in \cite{Oliva:2024rex}. Moreover, it would be interesting to match a quark-gluon plasma stage after the glasma phase, the former being modeled e.g. via relativistic transport theory. This would allow a more direct
comparison with experimental observables. These improvements will be presented in the near future.

\subsection*{Data availability}
The data that support the findings of this article are openly available in~\cite{data_folder_2505.08441}.

\subsection*{Acknowledgments}
M. R. acknowledges Francesco Acerbi and John Petrucci for inspiration.
The authors acknowledge numerous discussions with
Dana Avramescu, Bjorn Schenke and Raju Venugopalan.
This work has been partly funded by the
European Union – Next Generation EU through the
research grant number P2022Z4P4B “SOPHYA - Sustainable Optimised PHYsics Algorithms: fundamental
physics to build an advanced society” under the program
PRIN 2022 PNRR of the Italian Ministero dell’Università
e Ricerca (MUR).

\bibliography{biblio.bib}

\end{document}